\documentclass[conference]{IEEEtran}
\IEEEoverridecommandlockouts
\usepackage{cite}
\usepackage{amsmath,amssymb,amsfonts}
\usepackage{algorithmic}
\usepackage{graphicx}
\usepackage{textcomp}
\usepackage{xcolor}
\usepackage{tablefootnote}
\usepackage[numbers,sort&compress]{natbib}
\def\BibTeX{{\rm B\kern-.05em{\sc i\kern-.025em b}\kern-.08em
    T\kern-.1667em\lower.7ex\hbox{E}\kern-.125emX}}

\usepackage{booktabs}
\usepackage[flushleft]{threeparttable}
\usepackage{subcaption}

\begin{document}

\title{Fusion of Sentiment and Asset Price Predictions for Portfolio Optimization}

\author{\IEEEauthorblockN{Mufhumudzi Muthivhi}
\IEEEauthorblockA{\textit{Computer Science and Applied Mathematics} \\
\textit{University of the Witwatersrand}\\
Johannesburg, South Africa \\
1599695@students.wits.ac.za}
\and
\IEEEauthorblockN{Terence L. van Zyl}
\IEEEauthorblockA{\textit{Institute for Intelligent Systems} \\
\textit{University of Johannesburg}\\
Johannesburg, South Africa \\
tvanzyl@uj.ac.za}
}

\maketitle

\begin{abstract}
The fusion of public sentiment data in the form of text with stock price prediction is a topic of increasing interest within the financial community. However, the research literature seldom explores the application of investor sentiment in the Portfolio Selection problem. 
This paper aims to unpack and develop an enhanced understanding of the sentiment aware portfolio selection problem. 
To this end, the study uses a Semantic Attention Model to predict sentiment towards an asset. We select the optimal portfolio through a sentiment-aware Long Short Term Memory (LSTM) recurrent neural network for price prediction and a mean-variance strategy.
Our sentiment portfolio strategies achieved on average a significant increase in revenue above the non-sentiment aware models. However, the results show that our strategy does not outperform traditional portfolio allocation strategies from a stability perspective.
We argue that an improved fusion of sentiment prediction with a combination of price prediction and portfolio optimization leads to an enhanced portfolio selection strategy.
\end{abstract}

\begin{IEEEkeywords}
sentiment analysis, portfolio optimization, stock prediction, mean variance, long short term Memory, attention Model
\end{IEEEkeywords}

\section{Introduction}
Portfolio Optimization is an inherently complex Computational Finance problem. Progress often seems laboured despite substantial research made available. Portfolio selection is the process of finding the ``best'' allocation of wealth across a set of assets. Selection involves decisions around the return to risk ratio, dependent on the investors' financial goals. Researchers in the field have presented various methods to solve these challenges, the most established of which are Portfolio Optimization techniques~\cite{paskaramoorthy2020framework,perrin2020machine,van2021parden}.

Despite recent progress, the schools of behavioural finance and economics have criticized traditional models for taking a purely quantitative approach. Researchers continue to show that markets exist in significantly higher dimensional spaces and have many other unmeasured or quantified dependencies~\cite{nofsinger2005mood,smith2003economics}. The efficient market hypothesis is one such theory currently widely discussed at the forefront of economics and finance~\cite{fama1965behavior}. It emphasizes that future prices follow a random walk, and thus it is impossible to predict an asset's price. However, the theory is contentious and debate within academic circles continue to report on improved price prediction methods~\cite{patel2015predicting,fischer2018deep,long2019deep,nabipour2020deep}.
Recently, the fusion of investor sentiment as a feature in predictive models has become increasingly popular~\cite{ferreira2021Artificial}. Many articles report seeing better results by both late and early fusion of different models, i.e. price prediction, portfolio optimization and sentiment analysis.

This paper presents an investigation into an enhanced sentiment-aware portfolio selection strategy. The study begins by showing that the widely used sentiment analysis model, VADER, produces biased results. Thus the results motivate for the use of an improved sentiment model: Semantic Attention model to tackle the challenges of VADER~\cite{hutto2014vader}. Thereafter, we implemented a Long Short Term Memory (LSTM) price prediction model and a mean-variance strategy to obtain an optimal portfolio~\cite{laher2021deep}.
The presented sentiment-aware strategies outperformed traditional portfolio allocation strategies in an upward and downward trending market. However, our strategies were less robust than traditional approaches, resulting in increased exposure to risk. The study also explores various metrics to represent sentiment and implements methods to deal with biased labels due to erroneous classification of text data. The results compare two portfolio design techniques; one through a traditional mean-variance portfolio optimization strategy and another through a predictive LSTM framework, which predicts the best allocation of capital.

The study suggests that sentiment-aware portfolio strategies produce higher returns than traditional portfolio selection methods. The analysis details that price and investor sentiment is indeed correlated. Furthermore, the sentiment time-series can forecast returns and therefore, sentiment directly influences the movement of price. The results also indicate the significant presence of bias in widely used sentiment analysis tools.


\section{Background and Related Work}
\label{section:background}

Modern Portfolio Theory (MPT), pioneered by Harry Markowitz, sits at the centre of Computational Finance~\cite{markowitz1952Portfolio}. His work laid the foundation for research in 'Portfolio Selection'. MPT aims to optimize the allocation of wealth across $N$ risky assets. The Markowitz 'Mean-Variance Efficient Frontier' is used to select a diversified and more efficient portfolio. It allowed investors to select a portfolio based on their financial preference - maximising returns at a given risk level or minimising risk at a given expected return level. The 'Efficient Frontier' highlighted that a portfolio could achieve higher expected returns only through increased exposure to risk.

Earlier Portfolio Selection methods do not aim to predict future prices/trends based on historical data. Earlier methods aim to distribute capital across a set of assets efficiently. Traditional portfolio selection strategies include the Buy and Hold, Constant Rebalanced portfolios and Best stock strategy~\cite{Li2014survey}.

Researchers often compare the above traditional portfolio strategies to state of the art portfolio optimization methods such as the bio-objective mean-CVar model or the NSGA-II algorithm~\cite{mendoncca2020Multi-attribute}. The problem with portfolio optimization methods is that they are becoming increasingly complex to solve in polynomial time.
An alternative approach is to integrate price forecasting with portfolio optimization, vitalised by the recent appeal of artificial intelligence~\cite{ferreira2021Artificial}. A model is trained to predict price, volatility or trends using historical time series data, and then the information is used to enhance its portfolio selection process.

The "Efficient Market hypothesis" has been endorsed by many researchers since 1965~\cite{fama1965behavior}. It states that current stock prices reflect all historical information and only react to new information. Therefore stock prices follow a random walk. Similarly, Modern Portfolio Theory (MPT) takes a fully quantitative solution-driven approach and assumes an ideal market. However, an increasing number of recent literature has begun to dispute the hypothesis~\cite{butler1992Efficiency,kavussanos2001multivariate, gallagher2002macroeconomic, qian2007classifiers}.

Behavioural finance states that investors make decisions based on emotions and mood. Hence investors make irrational decisions, contrary to MPT. As a result, the market can not be "fully valid and efficient", as suggested by the "Efficient Market hypothesis". The market is affected by a concept known as \textit{Investor Sentiment} - which is symbolic of the public's mood/feeling towards an asset. The discovery has led to the inclusion of qualitative methods in financial models based on price prediction~\cite{huang2021Cryptocurrency,xing2019volatility}.

According to Ferreira \textit{et al.}~\cite{ferreira2021Artificial}, who performed a survey on state of the art Artificial Intelligence algorithms on stock market trading. A "combination of approaches indicates superiority over single methods" - this means that combining portfolio selection with price prediction and sentiment analysis always yields higher results than a single implementation.

Fusion of NLP and financial data is a non-trivial task as social media contains a significant number of sarcasm, metaphors and domain-specific terms. Both Bollen \textit{et al.}~\cite{bollen2011Twitter} and Koratamaddi \textit{et al.}~\cite{koratamaddi2021reinforcement} advocate for the use of more sophisticated NLP tools that can tackle the above problems. As a result, this paper proposes an enhanced sentiment analysis tool for a portfolio optimization problem. The paper makes use of a semantic Attention Model to predict the intensity of sentiment from text data~\cite{huang2019Image--text}.  Further, a sentiment-aware LSTM model was used to predict the price of an asset for each time period~\cite{malandri2018mood--driven}. We produce the optimal portfolio via a mean-variance portfolio optimization process~\cite{chen2021Mean--variance}. 

\begin{table}[htbp]
\caption{Sentiment labels for each asset}
\begin{center}
\begin{tabular}{l|rrr}
\toprule
\textbf{Asset} & \textbf{Positive}& \textbf{Negative} & \textbf{Neutral} \\
\bottomrule
\toprule
3M          &  54,896 & 19,276 &  74,338 \\
Microsoft   & 192,629 & 84,176 & 254,060 \\
Walt Disney & 113,489 & 48,420 & 147,167 \\
Verizon     & 100,511 & 15,767 &  30,491 \\
Cisco       & 182,881 & 44,558 &  30,416 \\
\bottomrule
\end{tabular}
\label{table:sentiment_count}
\end{center}
\end{table}

Finally, we also used another sentiment-aware LSTM early fusion model to predict the best allocation of capital and compare our results against the mean-variance strategy. Our intuition being that investor sentiment fusion with Portfolio Optimization should produce a superior portfolio strategy to traditional strategies.

\section{Data Analysis}
\label{section:data}

\subsection{Data Collection and Representation}

\subsubsection{Price Data}

The portfolio consisted of five Dow Jones Industrial Average (DJIA) companies from different industries. The companies have a long-standing reputation with their investors while still yielding sound returns on average. Therefore a favourable stock performance coupled with positive investor sentiment should generate a profitable portfolio.
We collected the price data from Yahoo Finance~\cite{Yahoowebsite}. The daily close adjusted price data ranges from 1 January 2001 to 31 December 2018. The period is rich with strong bull and bear market trends. 
The data was split into 70\% training data, 20\% testing data and 10\% validation data. The Training set ranges from 1 January 2001 to 11 December 2013. The Validation set; 12 December 2013 to 20 May 2015. The Testing set; 21 May 2015 to 31 December 2018.

\subsubsection{Sentiment Data}

We scraped the sentiment data from the Google and Twitter social media platforms. Google News Articles aggregates articles written by prominent news outlets. We used Selenium WebDriver to scrape news articles~\cite{Seleniumwebsite} 
We scraped news articles relevant to the five companies based on the unique name of the stock and its ticker symbol. The date and topic of the articles were collected.  snscrape was used as a scraper for Twitter tweets~\cite{snscrapewebsite}. 
The train, test and validation split, for the sentiment data, was divided similarly to the splits done on the price data. 

\subsubsection{Sentiment Labeling}

Unlike other classification problems, our data was not labelled, i.e. we were not aware of an article's or tweet's positive or negative sentiment. Most research makes use of VADER - "a lexicon and rule-based sentiment analysis tool" specifically trained on Twitter tweets~\cite{hutto2014vader}. VADER can interpret slang words, emoticons, emojis, contractions, and other natural language attributes. VADER uses a dictionary that maps lexical features to emotion intensities that quantify sentiment as positive or negative. The intensities are known as polarity scores.

\begin{figure}[htbp]
\begin{subfigure}[b]{\columnwidth}
\includegraphics[width=\columnwidth]{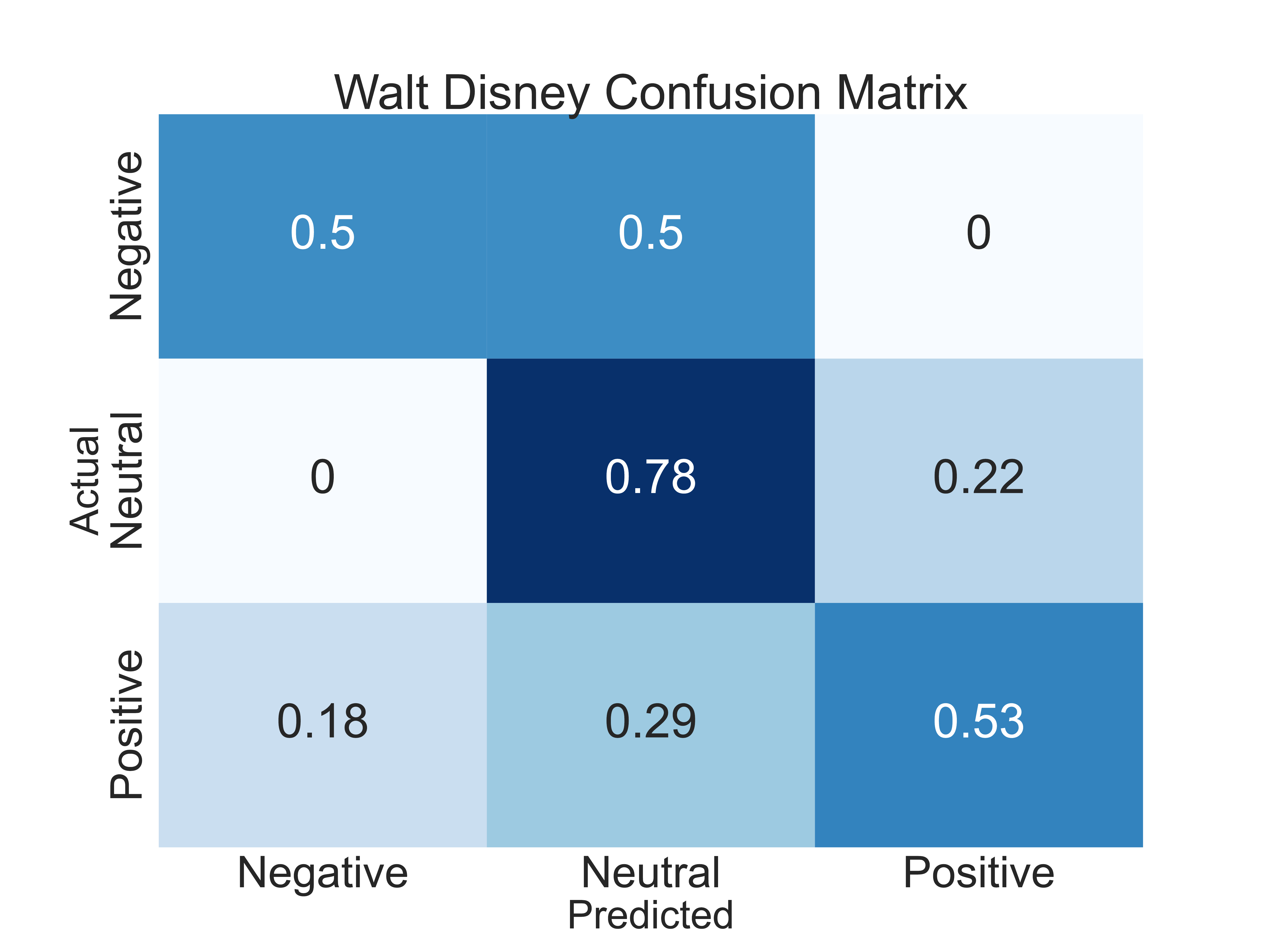}
\end{subfigure}
\begin{subfigure}[b]{\columnwidth}
\includegraphics[width=\columnwidth]{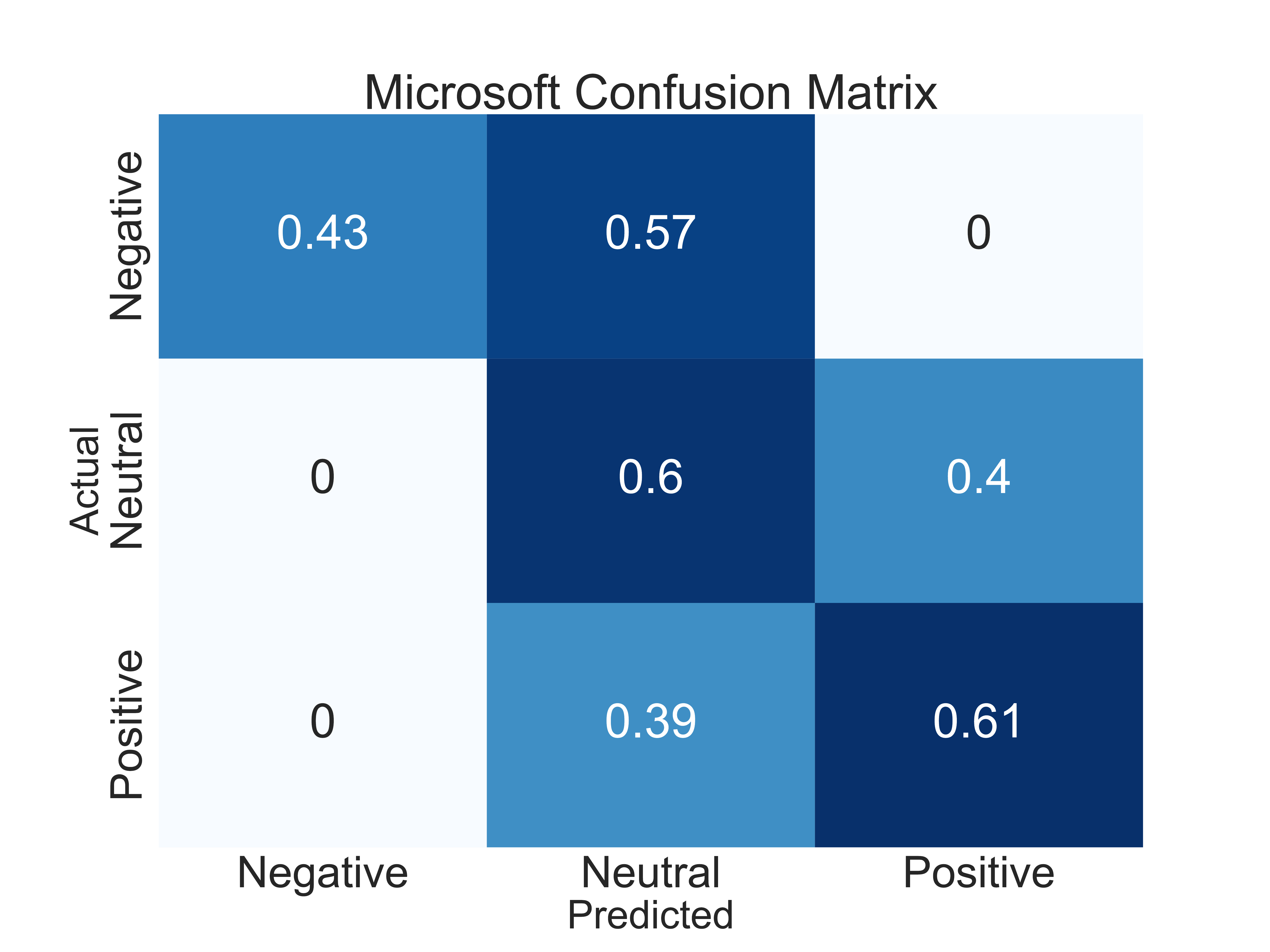}
\end{subfigure}
\caption{Confusion matrix of sampled data for the Walt Disney and Microsoft text data. Each row in the confusion matrix sums up to one.}
\label{figure:confusion}
\end{figure}

There are three possible labels; \textit{Positive}, \textit{Negative} and \textit{Neutral} sentiment for each text $t$. We retrieved the polarity scores of Google News Articles with VADER. Table~\ref{table:sentiment_count} depicts the Positive, Negative and Neutral sentiment counts for each company.

\subsubsection{Quality of Sentiment Data}

\begin{figure}[htbp]
\begin{subfigure}[b]{\columnwidth}
\centerline{\includegraphics[width=\columnwidth]{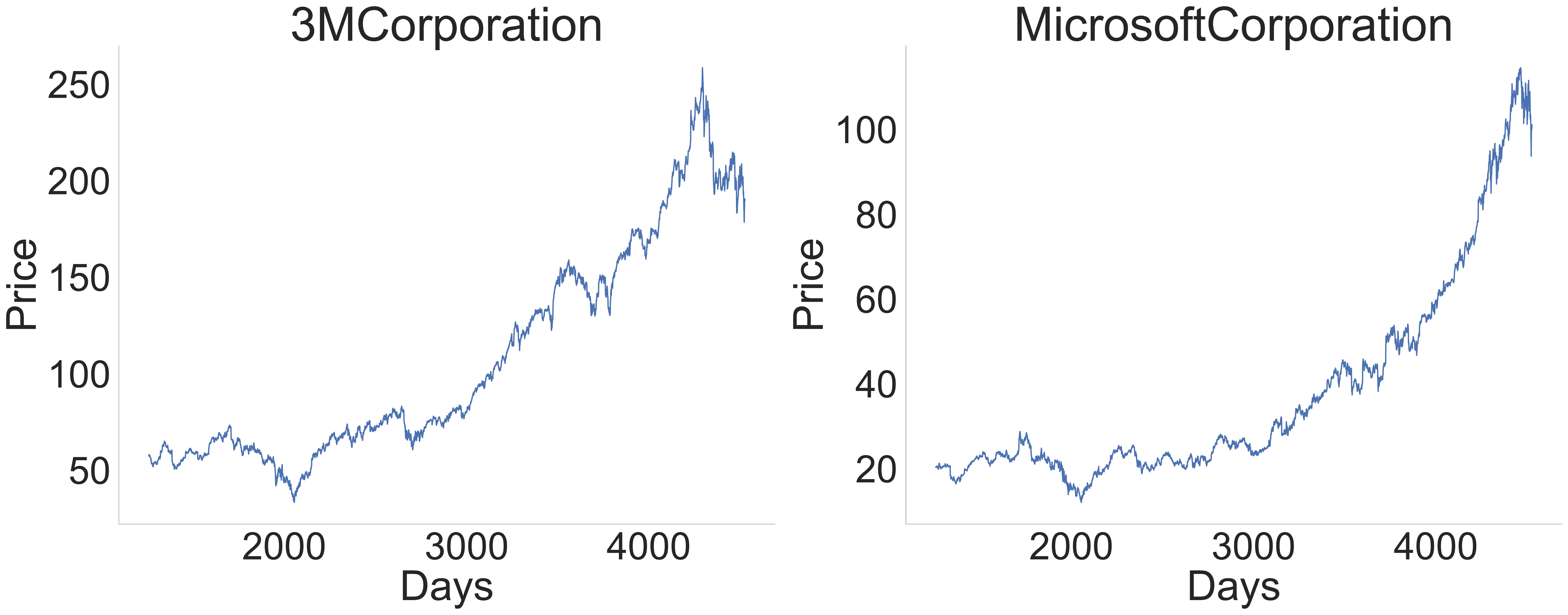}}
\end{subfigure}
\begin{subfigure}[b]{\columnwidth}
\centerline{\includegraphics[width=\columnwidth]{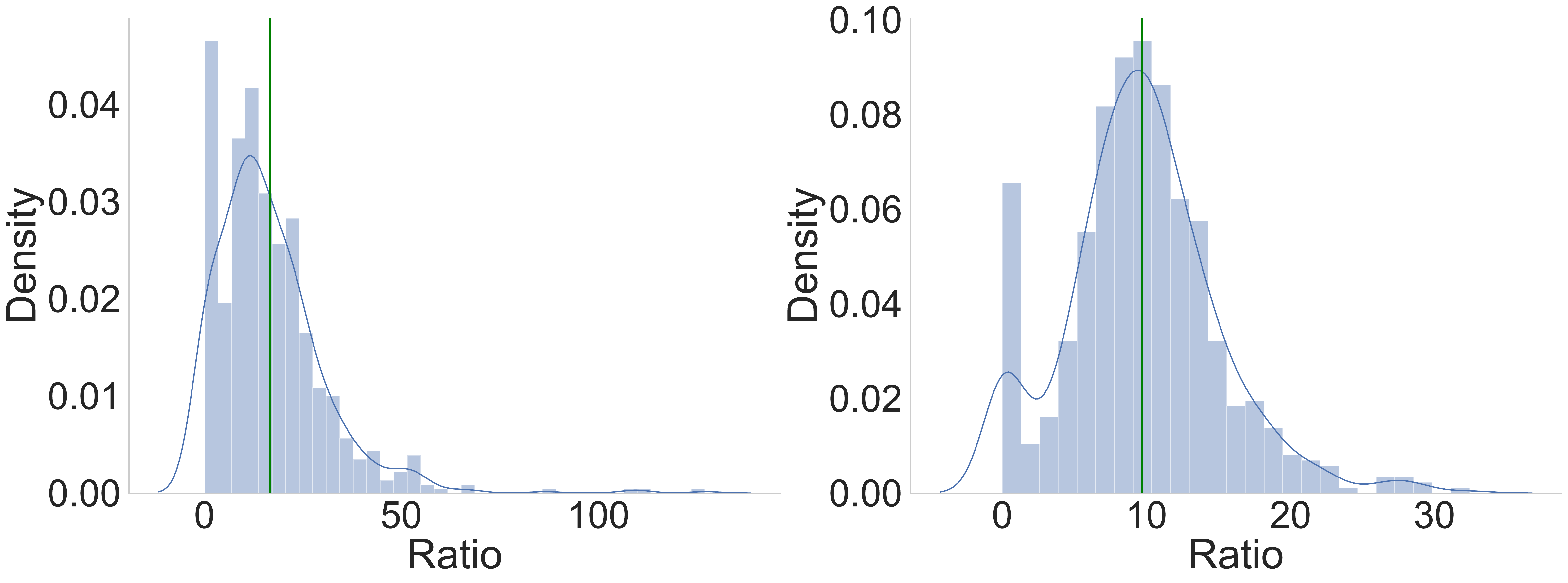}}
\end{subfigure}
\caption{Price charts for two assets against the distribution of the ratio of weekly sentiment labels}
\label{figure:ratio_up}
\end{figure}

Does the sentiment data accurately describe the investor's mood? 
According to Table~\ref{table:sentiment_count}, VADER had repeatedly assigned most text as \textit{Neutral} sentiment. Meaning that VADER gave the text a polarity score of zero. Either VADER is biased towards:
\begin{enumerate}
\item the majority of the text as having no positive or negative emotions; or
\item it was not able to derive the emotion from the text.
\end{enumerate}
The study investigates these issues by taking a stratified sample of the data to examine VADER's performance. That is, each sample was representative of the population's \textit{Positive}, \textit{Negative} and \textit{Neutral} sentiment proportions for each asset Table~\ref{table:sentiment_count}. We manually labelled 30 random samples from the Microsoft and Walt Disney datasets as Positive, Negative or Neutral sentiment. Then passed the examples to VADER and assessed the accuracy of its predictions.

About 43\% and 60\% of the examples were labelled accurately for the Microsoft and Walt Disney text data, respectively. The accuracy scores were not remarkably high. However, it is above 33\%, the average performance for a three-class classification problem. Again, VADER tends to label most of the data as Neutral.

We present the results using a confusion matrix~\ref{figure:confusion}. VADER had predicted the true \textit{Positive} and \textit{Negative} labels as \textit{Neutral} sentiment and the \textit{Neutral} label as \textit{Positive} sentiment. For instance, take the \textit{Walt Disney} confusion matrix - 53\% of the \textit{Positive} labels were labeled accurately and 29\% labeled incorrectly as \textit{Neutral} and the remaining 18\% were labeled as \textit{Negative} sentiment. A similar case exists for the \textit{Negative} labels. 50\% of the \textit{Negative} labels were labeled correctly and the remaining 50\% was predicted as \textit{Neutral} sentiment. This means that VADER fails to capture the negative emotions that exist in text and instead defaults to \textit{Neutral}.

Furthermore, the heat map of the confusion matrix was warmer towards the bottom right, more specifically,

\begin{figure}[htbp]
\begin{subfigure}[b]{\columnwidth}
\includegraphics[width=\columnwidth]{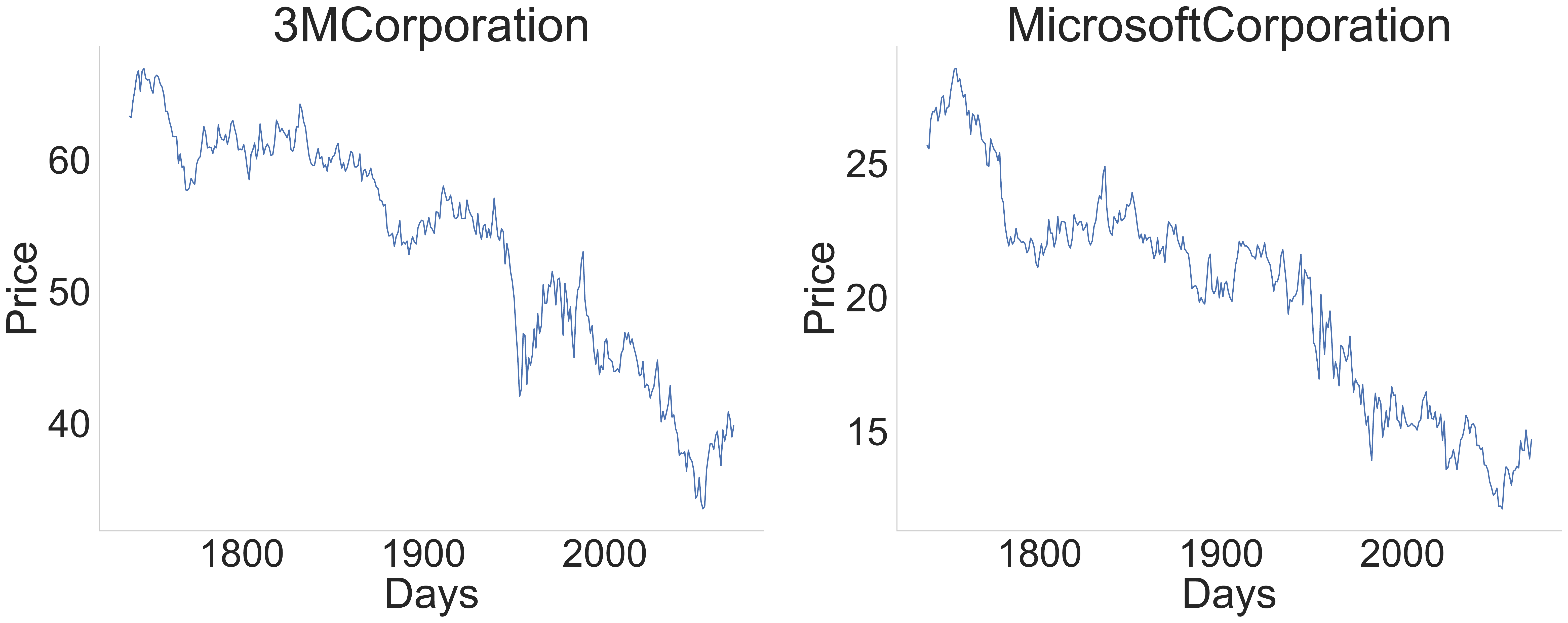}
\end{subfigure}
\begin{subfigure}[b]{\columnwidth}
\includegraphics[width=\columnwidth]{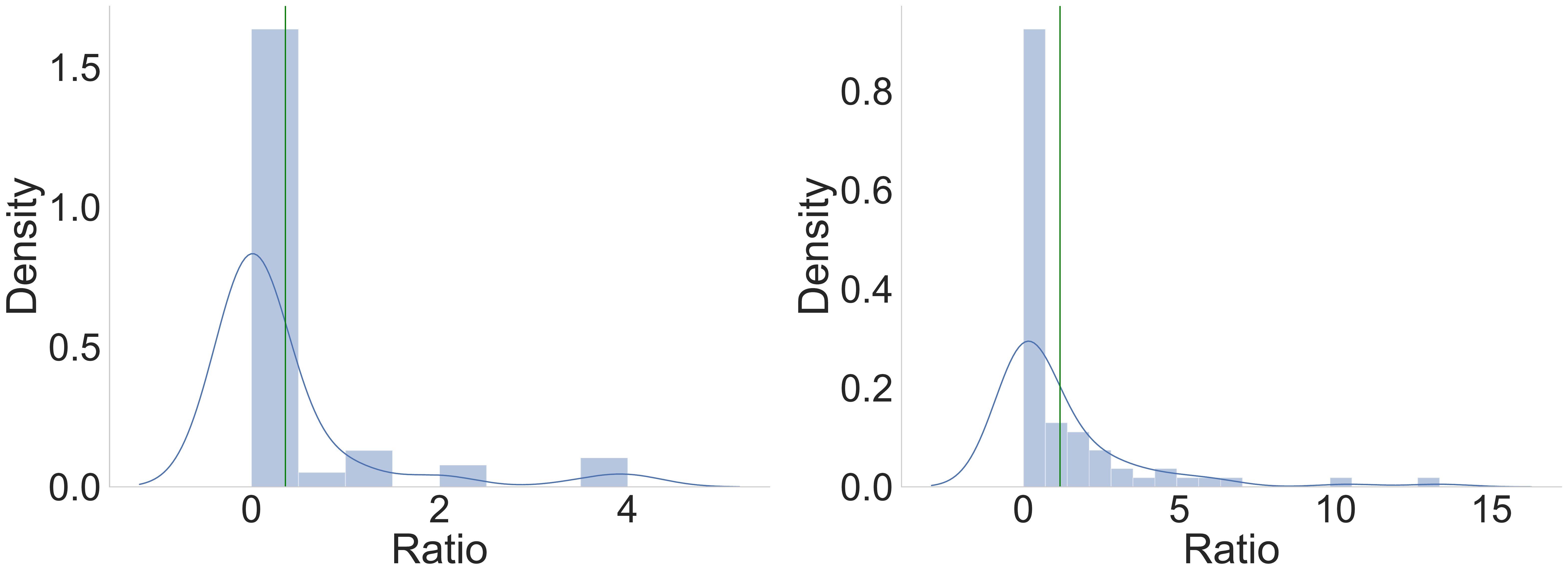}
\end{subfigure}
\caption{Great Recession price charts for two assets against the distribution of the ratio of weekly sentiment labels}
\label{figure:ratio_down}
\end{figure}

towards the \textit{Positive} label. It may also be the case that VADER was in favour of selecting \textit{Positive} labels but defaults to \textit{Neutral}. A similar case exists with the Microsoft confusion matrix~\ref{figure:confusion}. Nonetheless, we achieved an above 33\% accuracy score on a three-class classification model. This accuracy score means our model accurately describes investor sentiment half the number of times.

\subsubsection{Daily and Weekly Sentiment Data}
The general trend amongst researchers is to make daily predictions of price or stock allocations. However, some days may have more sentiment data points than others. The daily sentiment count trends upwards every year as Twitter and Google become more popular with more content generated on the platforms.
For instance, 85\% of \textit{3M Corporation's} total number of days - had examples of less than 30. Our objective was to observe enough sentiment per window so that we have a good perception of the entire window's sentiment. Meaning, a single investor's sentiment within one window would not be a representative sample of the entire market's mood towards an asset. We selected a lower bound of 30 data points for each window. 
Then made use of a weekly window and aggregate the entire windows sentiment of an asset into one weekly metric. As a result, we observed an improvement in the number of sentiment data-points per week, \textit{3M Corporation} now had 76\% of its days, with sentiment counts greater than 30.

\section{Experimental Methodology}

\label{section:experimental}

\subsection{Correlation of Sentiment and Returns data}
The previous section described that a weekly window contains more information about sentiment than a daily window. Therefore, we need a statistic that measures the intensity of the market's mood within a seven-day window.
The study represents the intensity of weekly sentiment in several ways. Our objective is to capture the true nature of the market's mood towards an asset. This was achieved by selecting the statistic that best describes the correlation between returns and sentiment. Furthermore, the correlation between returns and sentiment must be statically significant.

There exists four test cases: 1) mean polarity score for the week 2) max polarity score for the week, 3) median polarity score for the week and 4) ratio of positive and negative counts.
First, consider the average weekly sentiment by taking the mean of the polarity scores in a seven-day window. The results showed a slight positive correlation between the average polarity scores and returns data, Table~\ref{table:correlation}. We measured the significance of observing what we saw in the sample data over the entire population. Then Rejected the null hypothesis at a p-value of 0.05, i.e. there was a significant correlation between returns and sentiment data in the entire sentiment and stock market price population.

We benchmark the results against the average weekly polarity scores used previously. Essentially, for the max and median polarity scores, the correlation scores were weaker, Table~\ref{table:correlation}. The max and median statistics emphasised positive sentiment labels, thus perpetuating the bias that already exists in the data.

\begin{table}[htbp]
\caption{Correlation of sentiment and returns data for each asset}
\begin{center}
\begin{tabular}{l|rrrrr}
\toprule
\textbf{Asset} & \textbf{mean} & \textbf{max} & \textbf{median} & \textbf{ratio}\\
\bottomrule
\toprule
3M & 0.12 & 0.08 & 0.14 & 0.09 \\

Microsoft & 0.18 & 0.08 & 0.12 & 0.13 \\

Disney & 0.17 & 0.08 & 0.16 & 0.11 \\

Nike & 0.05 & 0.07 & 0.06 & 0.01 \\

Walmart & 0.14 & 0.01 & 0.14 & 0.14 \\

\bottomrule
\end{tabular}
\label{table:correlation}
\end{center}
\end{table}

We formulated the positive and negative sentiment ratio by taking the total count of positive labels for the week divided by the total count of negative labels for the week. For instance, a ratio of three implies that we have three positive sentiment labels for every negative sentiment label for that week. The distribution plots in Fig.~\ref{figure:ratio_up} convey a higher response to bullish and bearish events. The average ratio sat above 10 for each stock during a bullish event and dropped significantly closer to zero during a bearish event, Fig.~\ref{figure:ratio_down}. The correlation score between returns and sentiment did not improve significantly compared to the average weekly sentiment, Table~\ref{table:correlation}. However, the price and sentiment correlation score had a slight improvement. As a result, we used the price data for our LSTM model instead of the returns data. The returns data requires a transformation on our original close adjusted price and thus a loss of information.

\subsection{Causation}

The following results aim to show that the sentiment time series can forecast returns. Proving correlation does not imply causation of the two-time series. Hence, we investigated methods to prove that the lagged sentiment values can predict returns. We used the Granger’s Causality Test, similar to the implementation done by Bollen \textit{et al.}~\cite{bollen2011Twitter}.

The Bivariate Granger causality analysis test can explore the interdependence between two-time series. The test assumes that if a variable $X$ causes $Y$, then changes in $X$ systematically occur before changes in $Y$~\cite{bollen2011Twitter}. Our objective was to test for causation by observing if one time series had some predictive information over the other, i.e. do past values of $X$ aid in predicting $y_t$ in time $t$~\cite{gilbert2010widespread}.
Consider two linear regression equations:
\begin{equation}
    L_1: R_t = \alpha + \sum_{i=1}^n \beta_i R_{t-i} + \epsilon_t\label{equation:L1}
\end{equation}
\begin{equation}
    L_2: R_t = \alpha + \sum_{i=1}^n \beta_i R_{t-i} + \sum_{i=1}^n \gamma_i S_{t-i} + \epsilon_t\label{equation:L2}
\end{equation}
where $L_1$ uses $n$ lagged values of the returns $R$ to predict returns at time $t$. $L_2$ uses the $n$ lagged sentiment values plus returns to predict the returns for time $t$. To test for causality, we compared the variance explained by $L_1$ and $L_2$. 
Set the null hypothesis as \textit{the coefficients of past values in the regression equation is zero} and choose to reject the null hypothesis at a significance level of 0.05.

\begin{table}[htbp]
\caption{p-values of bivariate Granger-causality}
\begin{center}
\begin{tabular}{l|rrrrr}
\toprule
\textbf{Lags} & \textbf{3M} & \textbf{Microsoft} & \textbf{Disney} & \textbf{Nike} & \textbf{Walmart} \\
\bottomrule
\toprule
L1 & 0.635300 & 0.397800 & 0.739000 & \textbf{0.018800} & 0.051700 \\

L2 & 0.195000 & 0.569100 & 0.452800 & \textbf{0.042100} & 0.064800 \\

L3 & 0.327400 & 0.559500 & 0.627600 & \textbf{0.019300} & 0.130300 \\

L4 & 0.475000 & 0.440300 & 0.654700 & \textbf{0.049900} & 0.207000 \\

L5 & 0.651000 & 0.498400 & 0.096600 & 0.076100 & 0.290300 \\

L6 & 0.691500 & 0.317500 & 0.088800 & \textbf{0.013400} & 0.529600 \\

L7 & 0.776400 & 0.400000 & 0.125200 & 0.065300 & 0.739700 \\

L8 & 0.494100 & \textbf{0.046000} & \textbf{0.042100} & 0.087600 & 0.673900 \\
\bottomrule
\end{tabular}
\label{table:granger}
\end{center}
\end{table}


The results in Table~\ref{table:granger} presents the p-values achieved by each asset up to eight lags. We observed that \textit{Microsoft} and \textit{Walt Disney} are significant at lag 8 whereas \textit{Nike} at lag 1. \textit{3M} and \textit{Walmart} have significant lag values at values beyond eight. 

Given that we had some interdependence between the sentiment and returns time series, we can now use a fusion of sentiment and returns data to make forecasts. Thus, our sentiment time series had some predictive information over the returns time series.

\subsection{Models}

The paper used a Semantic Attention Model, similar to the implementation done by Huang \textit{et al.} \cite{huang2019Image--text}, to predict sentiment labels from text data. 
Further, a sentiment aware LSTM model is used to predict price. Then the predicted price is used to find the optimal portfolio based on a mean-variance framework. Finally, we discuss the implementation of a few benchmarks for comparative analysis.


\subsubsection{Semantic Attention Model}


Huang \textit{et al.}~\cite{huang2019Image--text} use a similar Semantic Attention text sentiment analysis model implemented by Wang \textit{et al.} and Chen \textit{et al.}~\cite{wang2016attention, chen2016neural}. However, different in its approach in highlighting the most important words due to its end-to-end process. 

The Semantic Attention model can focus on words in the text that are more important than others and capture the most discriminative features. Its objective is to emphasize the most emotive words.


The model makes use of an Adam replacement optimization algorithm to learn the parameters by minimizing the loss obtained from a \textit{Multi Label Soft Margin} loss function.

The model hyperparameters were tuned using Ray Tune, an industry-standard scalable hyperparameter tuning library~\cite{liaw2018tune}. The hyperparameters consist of the learning rate, batch size and the sentence length of our text, i.e. the number of encoded words in the text.

Each text was padded and truncated to a fixed size of 11, 21 and 31. 11 is the length of the encoded text below one standard deviation of the average length of all texts. Whereas 21 is above one standard deviation. Furthermore, 31 is three standard deviations above the average length of all texts.

The results chose a learning rate of 0.002, a batch size of 32 and a sentence length of 21.

\subsubsection{Price Prediction LSTM model}


LSTMs belong to the Recurrent Neural Networks (RNN) family with the additional feature of learning long time dependencies. The LSTM does this by using so-called forget and remember gates. These gates dictate which information is essential to propagate throughout the model and which to dismiss. 


The LSTM model must output the predicted price  $p_i^t$ for asset $i$ in time $t$. As previously determined, we looked at a seven-day window of our time series data. Meaning, the LSTM model reads in data from six days ago and predicts the price for the seventh day at each iteration. The input variables consist of the close adjusted stock price, total likes, total retweets, total comments, volume, and the ratio of positive to negative sentiment per day. Our portfolio consists of five stocks; as a result, there is a total of 30 predictor variables. 

The study used the validation data to measure the performance of the LSTM model. Our objective was to decrease the Mean Squared Error (MSE) loss between the predicted price and the actual price of an asset. The model made use of an \textit{Adam} replacement optimization algorithm and a \textit{MSE} loss function. Again, model hyperparameters are tuned using raytune~\cite{liaw2018tune}. The results chose a learning rate of 0.004, 3 layers and 13 hidden neurons.

\begin{figure*}[htbp]
\centerline{\includegraphics[scale=0.25]{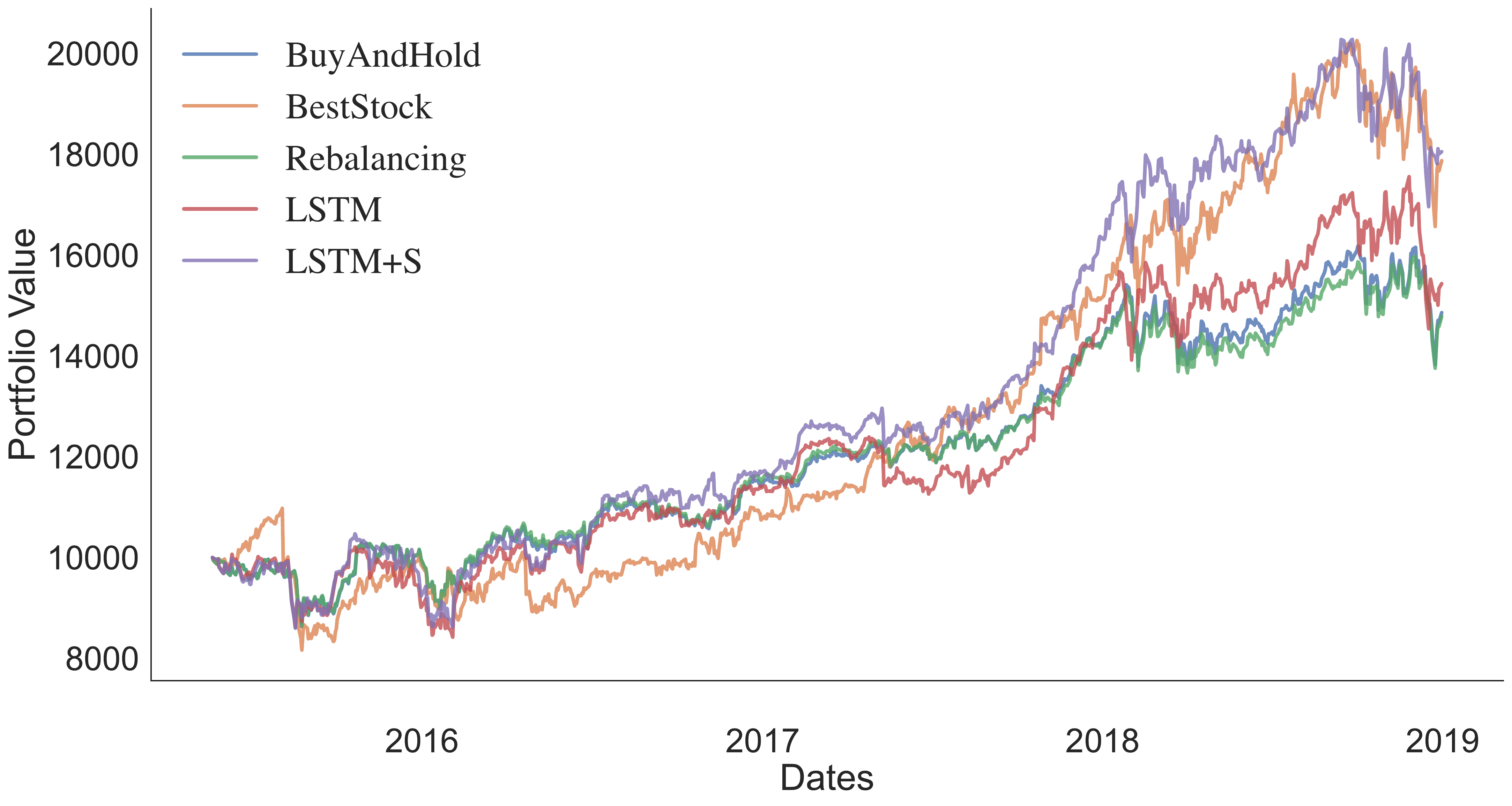}}
\caption{Comparison of daily portfolio values achieved by the portfolio strategies}
\label{figure:up_market}
\end{figure*}

\subsubsection{Mean-variance model}


The mean-variance model, of  Markowitz, is aimed at a trade-off between maximum returns and minimum risk~\cite{markowitz1952Portfolio}. It assumes that an investor will seek the highest returns:
\begin{equation}
    \max \sum_{i=1}^N x_i \mu_i \label{equation:S13}
\end{equation}
possible whilst pursuing the lowest risk possible:
\begin{equation}
    \text{s.t.}   \begin{cases}
                            \sum_{i=1}^N x_i = 1 \\
                            x_i \geq 0, \forall i = 1,...,n
                        \end{cases} \label{equation:S14}
\end{equation}
within a portfolio of $N$ risky assets. Where $\mu_i$ is the expected returns of asset $i$.


The mean-variance model is used to find the optimal portfolio at time $t$. The study used the Monte Carlo method to generate 50,000 portfolios of random weights. Then get the mean return and variance for each portfolio. Statistically, 50,000 random portfolios cover the majority of possible portfolios~\cite{wang2020preselection}. The aim was to select a portfolio from the 50,000 portfolios that sits along the Efficient frontier~\cite{markowitz1952Portfolio}. The study made use of the Sharpe Ratio $SR$ to select the optimal portfolio from the Efficient Frontier~\cite{sharpe1994ratio,Sharpe1964equilibrium}.

\subsubsection{Benchmarks}

The final portfolio optimization model consists of the Semantic Attention model, the LSTM price prediction model, and the mean-variance portfolio optimization strategy. We refer to our final portfolio optimization model as $LSTM+S$. 'S' denotes the sentiment aware LSTM model with a mean-variance portfolio optimization strategy. The results of our final model ($LSTM+S$) are first compared to a non-sentiment aware LSTM model that also made use of a mean-variance portfolio optimization strategy. We refer to the non-sentiment aware LSTM model as $LSTM$. In addition, the paper has assessed the statistical significance of the $LSTM+S$ portfolio performance. Then, the study contrasts the optimal portfolios produced by the mean-variance strategy to three other portfolio selection benchmarks. First, a uniform Buy and Hold portfolio ($BAH$) which assigns equal weights to each asset for the entire period $T$. Secondly, a uniform Constant Rebalancing Portfolio ($Rebalancing$) which reassigns equal allocations of capital at each time interval $t$. Finally, a Best Stock ($BestStock$) portfolio strategy allocates all the capital to the best performing asset over the initial investment period to the current day in observation.~\cite{Li2014survey, kelly1956interpretation,cover1991Universal}. 

\subsubsection{Performance Measures}

We consider the fAPV, the Accumulated Portfolio value at the end of the period divided by the initial capital. To measure the risk factors of a portfolio, we applied the Sharpe Ratio ($SR$) over the entire period:
\begin{equation}
    SR = \frac{\mathbb{E}_t[R_p / R_{_\text{BH}}]}{\sigma(R_p) / \sigma(R_{_\text{BH}})} \label{equation:S15}
\end{equation}
which is the risk-adjusted return. $R_p$ are the returns achieved by portfolio $p$, and $R_{_\text{BH}}$ are the returns of the Buy and Hold portfolio (BH). We set the BH as the base value such that the $SR$ for BH is equal to 1.0.

One issue with the Sharpe Ratio is that it treats both upward and downward volatility as equivalent. The Maximum Drawdown ($MDD$) associates upwards volatility to positive returns and downward volatility to negative returns~\cite{magdon2004maximum}. Mathematically:
\begin{equation}
    MDD = \max_{0 < t < T} \frac{\text{Value}_t - \text{Value}_T}{\text{Value}_t} \label{equation:S16}
\end{equation}
where $\text{Value}_t$ is the capital value of our portfolio at time $t = \{1, 2,..., T\}$. A portfolio allocation strategy with a large $MDD$ will result in investor panic and thus pressure to withdraw.

Finally, we calculate the annualized return by measuring the portfolio's improvement over a year.

\section{Results}
\label{section:results}

Our primary benchmark is the Buy and Hold portfolio since its daily returns correspond to the average daily returns of the assets in the portfolio. This strategy performs better than value and price-weighted portfolios in terms of mean returns and Sharpe Ratio. However, it usually bears higher risk~\cite{plyakha2012does}. 
We compute the returns of a portfolio strategy against the Buy and Hold benchmark and refer to this as the benchmark value (BV). Mathematically, divide the final portfolio value of a portfolio strategy to the final value of the Buy and Hold portfolio.

To get the wealth achieved by a portfolio strategy for each time $t$:
\begin{equation}
    w_t = w_{t-1} \sum_{n=1}^N r_{n}^t s_{n}^t \label{equation:S16}
\end{equation}
where $w_t$ is the wealth at time $t$, $r_{n}^t$ are the returns of asset $n$ and $s_{n}^t$ is the weight assigned to that asset.


This study produced the results for two different scenarios. Table~\ref{table:top3_up} depicts the daily capital achieved from the period May 2015 to Dec 2018, and Table~\ref{table:top3_down} depicts the results of the portfolio strategies on a down market. The study defines a down market as the period where all assets experience a downwards trend in asset value. We chose the period 13 Dec 2014 to 17 Sep 2015. All of the assets experienced a bearish movement in price. Furthermore, enough sentiment data was available for analysis during the observed time period. 

This paper has designed a portfolio prediction sentiment aware LSTM model ($LSTM+S$) that outputs the optimal portfolio through a mean-variance framework.
The study compares three portfolio allocation benchmarks to the ($LSTM+S$): Buy and Hold ($BuyHold$), Best Stock ($BestStock$), and  the uniform  Constant Rebalancing portfolio ($Rebalancing$). Finally, we compare our results to a non-sentiment aware LSTM model ($LSTM$).

The results show that the $LSTM+S$ outperformed all strategies, Fig.~\ref{figure:up_market}, however, performed second best during a down market Table.~\ref{table:top3_down}. 
Both LSTM based models outperformed all portfolio allocation benchmarks. The $LSTM$ model performed best during a down market, with $LSTM+S$ a close second. 

\begin{table}[htbp]
\caption{Performance metrics with Top 3 in bold}
\begin{center}
\resizebox{\columnwidth}{!}{%
\begin{tabular}{l|rrrrrr}
\toprule
 \textbf{Models} & \textbf{Capital} & \textbf{fAPV} & \textbf{BV} & \textbf{SR} & \textbf{MDD(\%)} & \textbf{AR(\%)} \\
\bottomrule
\toprule
Buy and Hold & 14856.68 & 1.49 & 1.00 & 1.00 & \textbf{8.15} & 17.16\\

Best Stock & \textbf{17871.20} & 1.79 & \textbf{1.20} & \textbf{4.17} & 11.76 & \textbf{26.14}\\

Rebalancing & 14781.47 & 1.48 & 0.99 & 0.91 & \textbf{7.39} & 16.92\\

LSTM & \textbf{16966.97} & 1.70 & \textbf{1.14} & \textbf{3.26} & 12.98 & \textbf{23.55}\\

LSTM Sentiment & \textbf{18047.80} & 1.80 & \textbf{1.21} & \textbf{1.17} & \textbf{10.99} & \textbf{26.64}\\
\bottomrule
\end{tabular}
\label{table:top3_up}
}
\end{center}
\end{table}

\begin{table}[htbp]
\caption{Performance metrics (for a down market) with Top 3 in bold}
\begin{center}
\resizebox{\columnwidth}{!}{%
\begin{tabular}{l|rrrrrr}
\toprule
 \textbf{Models} & \textbf{Capital} & \textbf{fAPV} & \textbf{BV} & \textbf{SR} & \textbf{MDD(\%)} & \textbf{AR(\%)} \\
\bottomrule
\toprule
Buy and Hold & 	9766.90 & 0.98 & 1.00 & \textbf{5.27} & \textbf{8.56} & -0.94\\

Best Stock & \textbf{9972.24} & 0.99 & \textbf{1.02} & \textbf{7.68} & 14.37 & \textbf{-0.11}\\

Rebalancing & 9809.24 & 0.98 & 1.00 & \textbf{5.39} & \textbf{8.44} & -0.77\\

LSTM  & \textbf{10978.207} & 1.10 & \textbf{1.12} & -14.57 & \textbf{8.81} & \textbf{3.80}\\

LSTM Sentiment & \textbf{10702.91} & 1.07 & \textbf{1.10} & -2.01 & 11.18 & \textbf{2.75}\\
\bottomrule
\end{tabular}
\label{table:top3_down}
}
\end{center}
\end{table}

\begin{table}[htbp]
\caption{Paired t-test of returns produced by LSTM with and without sentiment}
\begin{center}
\resizebox{.3\textwidth}{!}{%
\begin{tabular}{l|rr}
\toprule
 \textbf{Models} & \textbf{LSTM+S} & \textbf{LSTM} \\
\bottomrule
\toprule
Mean & 	7106.00 & 5311.55 \\
Standard deviation & 916.32 & 1042.63 \\
Observations & 10 & 10 \\
\midrule
degrees of freedom  & \multicolumn{2}{c}{9} \\
t statistic & \multicolumn{2}{c}{3.9507} \\
p-value & \multicolumn{2}{c}{0.0033} \\
\bottomrule
\end{tabular}
\label{table:significance}
}
\end{center}
\end{table}

This may have been due to the \textit{Positive} label bias, since we did not have enough \textit{Negative} labels to predict bearish events.


Table~\ref{table:top3_up}~and~\ref{table:top3_down} depict the performance measures of each portfolio strategy. As noted, the $LSTM+S$ achieved the highest accumulated capital of 18047.80. The $LSTM$ without sentiment achieved the highest accumulated capital in a down market, with the $LSTM+S$ performing slightly below. 
For both, Tables~\ref{table:top3_up} and~\ref{table:top3_down}, the annualized returns are also the highest in LSTM models that include sentiment data. 
However, the $LSTM+S$ was the most unstable with a low Sharpe Ratio and high MDD. Hence this strategy will not necessarily be the most popular amongst risk-averse investors. The benchmark portfolios, such as Buy and Hold and Rebalancing, provided for a less risky portfolio than the price prediction and mean-variance portfolios.

The $LSTM+S$ and $LSTM$ were the only portfolio strategies that produced a profit during a down market. The LSTM model's produce profitable portfolios even during strong bear markets. However, this was achieved through riskier allocations of capital.

The results show that adding sentiment to an LSTM and portfolio optimization model increases the portfolio returns.
To asses the statistical significance of this increment we perform a paired $t$ test on the returns of the LSTM model with and without sentiment, i.e. $LSTM+S$ and $LSTM$ respectively.

Table~\ref{table:significance} depicts a statistically significant mean difference between the LSTM portfolio returns with and without sentiment. We observe a p-value of 0.33\% which indicates that the sentiment data was informative in producing an optimal portfolio. 

\section{Conclusion}
\label{section:conclusion}
This paper aimed to design an optimal portfolio given the sentiment of a class of assets. We used an enhanced sentiment aware portfolio selection strategy, consisting of a Semantic Attention model, LSTM model and a mean-variance strategy.

The study discovered that the sentiment labels produced by VADER were significantly biased. The sentiment aware portfolio optimization strategy outperformed traditional portfolio strategies. However, our portfolio strategy was more unstable and less attractive to risk-averse investors.

This study suggests that sentiment aware portfolio strategies produce overall positive returns. The same applies to markets experiencing a downward trend. However, we achieve this superior performance through riskier allocations of capital.

Data collected from social media was not pre-labelled. Hence, we usually depend on unsupervised classification algorithms. A significant portion of investor sentiment analysis is domain-specific. Using pre-trained models on a unique dataset produces sub-optimal results. A superior approach would be to use datasets that the content publisher has labelled. Take, for instance, the customer reviews dataset.

One approach to improving sentiment labels' quality is integrating an image sentiment analysis tool such as a Visual Attention model. Currently, social media data is dependent on visual context. Initially, we had implemented such a model. However, various issues such as the size of RGB image pixels, images in the form of price charts and inappropriate images affected the progress.

\bibliographystyle{IEEEtran}
\bibliography{main}

\vspace{12pt}

\end{document}